\documentclass{emulateapj}
\usepackage{apjfonts}
\usepackage[]{natbib}
\usepackage{graphics}
\usepackage{color}
     
\newcommand{\etal}{et al.}  
\newcommand{\per}{\ensuremath{^{-1}}}

\newcommand{\msun}{\ensuremath{{M}_{\odot}}}
 
\newcommand{\kms}{km~s\ensuremath{^{-1}}}

\newcommand{\mbh}{\ensuremath{M_\mathrm{BH}}}
\newcommand{\chisq}{\ensuremath{\chi^2}}

\newcommand{\chisqdof}{\ensuremath{\chi^2_{\nu}}}

\newcommand{\msigma}{\ensuremath{\mbh-\sigma_{\star}}}

\newcommand{\mstar}{\ensuremath{M_\star}}

\newcommand{\sigmaturb}{\ensuremath{\sigma_\mathrm{turb}}}

\newcommand{\rg}{\ensuremath{r_\mathrm{g}}}
\newcommand{\vlos}{\ensuremath{v_\mathrm{LOS}}}
\newcommand{\sigmalos}{\ensuremath{\sigma_\mathrm{LOS}}}

\newcommand{\vsys}{\ensuremath{v_\mathrm{sys}}}

\newcommand{\vrot}{\ensuremath{v_\mathrm{rot}}}
\newcommand{\rfit}{\ensuremath{r_\mathrm{fit}}}

\begin{document} 

\title{Measurement of the Black Hole Mass in NGC 1332 from \\ ALMA
  Observations at 0.044 Arcsecond Resolution}

\author{
  Aaron J. Barth\altaffilmark{1},
  Benjamin D. Boizelle\altaffilmark{1},
  Jeremy Darling\altaffilmark{2},
  Andrew J. Baker\altaffilmark{3},  
 \\
  David A. Buote\altaffilmark{1}, 
  Luis C. Ho\altaffilmark{4},
  Jonelle L. Walsh\altaffilmark{5}}

\altaffiltext{1}{Department of Physics and Astronomy, 4129 Frederick
  Reines Hall, University of California, Irvine, CA, 92697-4575, USA;
  barth@uci.edu}

\altaffiltext{2}{Center for Astrophysics and Space Astronomy,
  Department of Astrophysical and Planetary Sciences, University of
  Colorado, 389 UCB, Boulder, CO 80309-0389, USA}

\altaffiltext{3}{Department of Physics and Astronomy, Rutgers, the
  State University of New Jersey, 136 Frelinghuysen Road, Piscataway,
  NJ 08854-8019, USA}

\altaffiltext{4}{Kavli Institute for Astronomy and Astrophysics,
  Peking University, Beijing 100871, China; Department of Astronomy,
  School of Physics, Peking University, Beijing 100871, China}

\altaffiltext{5}{George P. and Cynthia Woods Mitchell Institute for
  Fundamental Physics and Astronomy, Department of Physics and
  Astronomy, Texas A\&M University, College Station, TX 77843-4242,
  USA}

\begin{abstract}

  We present Atacama Large Millimeter/submillimeter Array (ALMA) Cycle
  3 observations of CO(2--1) emission from the circumnuclear disk in
  the E/S0 galaxy NGC 1332 at 0\farcs044 resolution. The disk exhibits
  regular rotational kinematics and central high-velocity emission
  ($\pm500$ \kms) consistent with the presence of a compact central
  mass. We construct models for a thin, dynamically cold disk in the
  gravitational potential of the host galaxy and black hole, and
  fit the beam-smeared model line profiles directly to the ALMA
  data cube.  Model fits successfully reproduce the disk kinematics
  out to $r=200$ pc. Fitting models just to spatial pixels within
  projected $r=50$ pc of the nucleus (two times larger than the black
  hole's gravitational radius of influence), we find
  $\mbh=(6.64_{-0.63}^{+0.65})\times10^8$ \msun.  This observation
  demonstrates ALMA's powerful capability to determine the masses of
  supermassive black holes by resolving gas kinematics on small
  angular scales in galaxy nuclei.

\end{abstract}


\section{Introduction}

The Atacama Large Millimeter/submillimeter Array (ALMA) has a
revolutionary capability to resolve cold molecular gas kinematics on
angular scales well below 1\arcsec. This opens the possibility of
using cold molecular gas disks as dynamical tracers to measure the
masses of supermassive black holes (BHs) in galaxy centers
\citep{davis2013b, onishi2015}.  Accurate measurement of a BH mass
(\mbh) generally requires observations that resolve kinematics within
the BH's gravitational radius of influence \rg, the radius within
which the BH dominates the gravitational potential, and ALMA is the
first mm/sub-mm interferometer having the capability to resolve
\rg\ in large numbers of galaxies.

NGC 1332 is a nearby S0 or E galaxy that contains a highly inclined
circumnuclear dust disk.  Its BH mass has been measured previously via
stellar dynamics \citep{rusli2011} and by modeling the hydrostatic
equilibrium of its X-ray emitting halo \citep{humphrey2009}, making it
an excellent laboratory for direct comparison of independent
measurement techniques.  In Paper I \citep{barth2016}, we described
0\farcs3-resolution Cycle 2 ALMA observations designed to test for the
presence of CO emission from within \rg. The data showed very clean
disklike rotation with a slight warp, and central emission extending
to $\pm500$ \kms. Our simplest dynamical model fits converged on
$\mbh=6.0\times10^8$ \msun, but beam smearing in the highly inclined
disk caused a severe degeneracy between rotation and turbulent motion
at small radii, precluding the derivation of stringent confidence
limits on \mbh. Here, we present new ALMA Cycle 3 observations at
0\farcs044 resolution and dynamical modeling of the data. We adopt a
distance of 22.3 Mpc to NGC 1332 for consistency with Paper I and with
\citet{rusli2011}; at this distance, 1\arcsec\ corresponds to 108.6
pc.

\section{Observations}

\begin{figure*}[t!]
  \scalebox{1.3}{\includegraphics{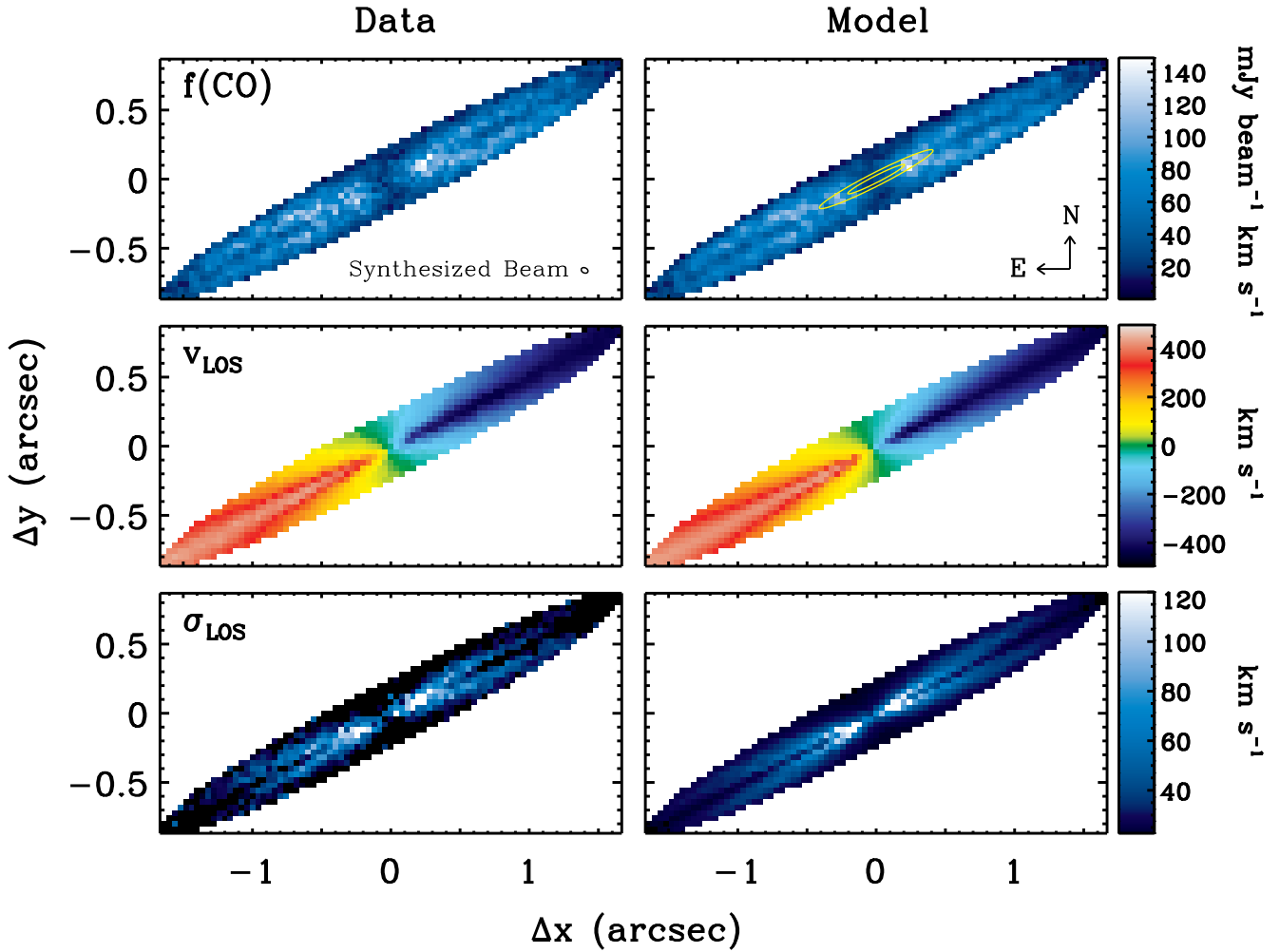}}
  \caption{Maps of CO(2--1) intensity, \vlos\ (relative to
    $\vsys=1562$ \kms), and \sigmalos\ in 0\farcs04 binned pixels for
    the data and best-fitting full-disk model. In the top right
    panel, the yellow ellipses denote the projected BH sphere of
    influence ($\rg=25$ pc), and the $\rfit=50$ pc region used for our
    final fits.}
  \label{fig:momentmaps}
\end{figure*}

NGC 1332 was observed in ALMA Band 6 as part of project
2015.1.00896.S. Three separate observations on 2015 September 16 and
23 were carried out, for a total on-source integration time of 101
min. The observations were obtained in a frequency band centered at
229.354 GHz, corresponding to the redshifted $^{12}$CO(2--1) line in
NGC 1332. We processed the data using the Common Astronomy Software
Applications \citep[CASA;][]{mcmullin2007} package to produce an
emission-line data cube with $0\farcs01$ pixels, a field of view of
$6\farcs4\times6\farcs4$, and 75 frequency channels with a channel
spacing of 15.4 MHz, corresponding to a velocity width of 20.1
\kms\ per channel for the CO(2--1) line.  The synthesized beam has
major and minor axis FWHM sizes of 0\farcs052 and 0\farcs037 with
major axis position angle (PA) 64\arcdeg, giving a geometric mean
resolution of 0\farcs044.

\begin{figure}
  \scalebox{0.8}{\includegraphics{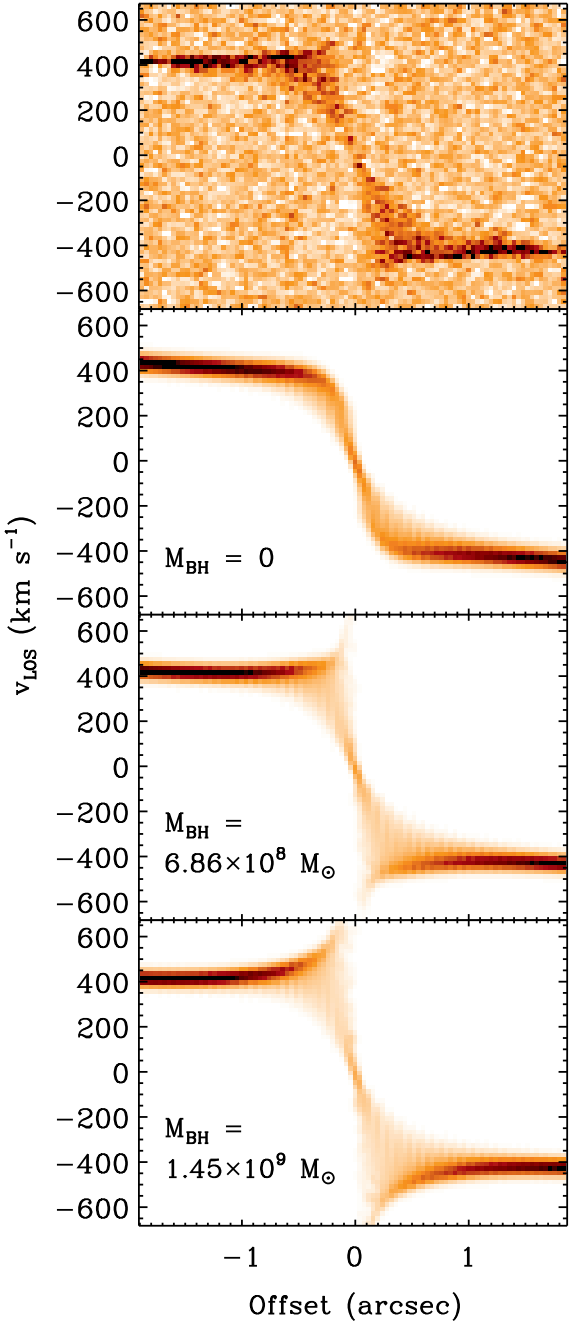}}
  \caption{PVDs generated by rotating the data cubes clockwise by
    27\arcdeg\ and extracting a single-pixel (0\farcs04) cut along the
    disk major axis. \emph{Top panel:} Observed PVD. \emph{Lower
      panels:} PVDs from full-disk model fits with \mbh\ fixed at $0$,
    $6.86\times10^8$ \msun\ (the best-fitting mass for the full-disk
    model fit), and $1.45\times10^9$ \msun\ (as measured by Rusli
    \etal\ 2011). The model PVDs are recomputed using a flat,
    noise-free CO surface brightness profile rather than the observed
    surface brightness profile, in order to illustrate the disk
    kinematics more clearly. The broad velocity spread at small radii
    is the result of beam smearing of unresolved rotation.}
  \label{fig:pvd}
\end{figure}

We created kinematic maps by fitting Gaussian line profiles to the
data. In order to increase the S/N of the profile fits, we first
binned the data cube spatially by block averaging in $4\times4$ pixel
blocks, producing 0\farcs04 pixels.  The CO intensity, line-of-sight
mean velocity (\vlos), and width (\sigmalos) are shown in Figure
\ref{fig:momentmaps}. The disk kinematics are illustrated over an
elliptical region with major and minor axes of 3\farcs7 and 0\farcs5,
corresponding to the largest region over which we were able to obtain
successful profile fits at each binned pixel. At the 0\farcs04 binned
pixel scale, the rms noise level is 0.142 mJy beam \per\ per channel,
and the the median S/N per 0\farcs04 spatial pixel is 7.2 over this
elliptical region.  The \vlos\ map illustrates remarkably regular
disklike rotation, with subtle signs of the mild kinematic twist or
warp we found in the Cycle 2 data. In the \sigmalos\ map, the ``X''
shaped feature is characteristic of rotational broadening in an
inclined disk. The major axis position-velocity diagram (PVD) exhibits
a central velocity upturn to $\approx500$ \kms\ (Figure
\ref{fig:pvd}), similar to that seen in the Cycle 2 data.

\section{Dynamical Modeling}

Our dynamical modeling follows the methods described in Paper I. We
model a thin disk in circular rotation in a gravitational potential
containing a central point mass and an extended mass distribution from
stars. The enclosed stellar mass $M_\star(r)$ is based on a
deprojected luminosity profile $L_\star(r)$ multiplied by a
mass-to-light ratio $\Upsilon$. We neglect the contributions of dark
matter and of molecular gas, both of which are $\lesssim10^8$
\msun\ within $r=250$\ pc; in contrast,
$M_\star(250~\mathrm{pc})\approx 10^{10}$ \msun\ (Paper I). Models are
computed on a spatial grid with 0\farcs01 pixels. At each grid
element, we determine the line-of-sight component of the disk's
rotation velocity (relative to the systemic velocity \vsys), observed
at inclination $i$ and major axis position angle $\Gamma$. The
emission line at each grid point is assumed to have a Gaussian
velocity profile, with turbulent velocity dispersion \sigmaturb.

This procedure produces a simulated cube with the same velocity
channel spacing as the data cube. The total line flux at each spatial
element is scaled to match a CO surface brightness map measured from
the ALMA data. Then, each velocity channel image in the model is
convolved with the ALMA synthesized beam, and the model is rebinned to
match the 0\farcs04 spatial scale of the binned ALMA data cube. The
goodness of fit (quantified by \chisq) is computed by direct
comparison of the model cube with the binned ALMA data. Calculating
\chisq\ at a spatial scale of approximately one pixel per synthesized
beam substantially mitigates the problem of spatially correlated noise
in the data, which would be a significant issue if \chisq\ were
computed for 0\farcs01 pixels (see Paper I). Free parameters include
\mbh, $\Upsilon$, \sigmaturb, $i$, $\Gamma$, $\vsys$, the dynamical
center position $(x_0,y_0)$, and an overall normalization factor $f_0$
applied to the CO surface brightness model. We assume a uniform value
of $\sigmaturb$ over the disk surface and discuss this assumption in
\S\ref{sec:turbulence}.  We optimize the model fits using the
\texttt{amoeba} downhill simplex method in IDL.

To model $M_\star(r)$, we use the deprojected stellar luminosity
profile from \citet{rusli2011}, based on a combination of $K$-band VLT
adaptive-optics data at small radii ($r<4\farcs5$) and $R$-band
imaging at larger radii.  Their best-fitting stellar-dynamical model
implies an $R$-band mass-to-light ratio of $\Upsilon_R=7.35$
(J. Thomas, private communication).  \citet{rusli2011} give full
details of the $L_\star(r)$ measurement, and Paper I presents a plot
of $M_\star(r)$.

\section{Modeling Results}
\label{sec:results}

\subsection{Fits to the full disk}

We first describe models fitted to the entire disk, over the
elliptical region illustrated in Figure \ref{fig:momentmaps}. This
region contains 840 binned (0\farcs04) spatial pixels. We compute
\chisq\ over 56 frequency channels, for a total of 47040 data
points. With all model parameters free, the best fit is found at
$\mbh=6.86\times10^8$ \msun, with $\chisq_\mathrm{min} = 55206.9$ or
$\chisqdof = 1.17$. Other parameters for the best-fitting model are
$\Upsilon_R = 7.53$, $i=84\fdg1$, $\Gamma = 117\fdg2$, and
$\sigmaturb=22.2$ \kms; these are similar to the best-fitting values
found in Paper I.  Figure \ref{fig:chisq} shows $\Delta\chisq = \chisq
- \chisq_\mathrm{min}$ as a function of \mbh. The 68.3\% confidence
region (corresponding to $\Delta\chisq=1$) is
$(6.86\pm0.14)\times10^8$ \msun.  Kinematic maps measured from the
best-fitting model (Figure \ref{fig:momentmaps}) match the structure
in the data closely. While the model is generally successful at
reproducing the disk kinematics, a result of $\chisqdof=1.17$ for
47032 degrees of freedom is formally unacceptable. We attribute this
to systematic inadequacies of the model, such as neglect of the disk's
slight warp, the assumption of a spatially constant value of
\sigmaturb, and noise in the CO surface brightness model.

The apparently high precision of this result is typical of
gas-dynamical models fitted to a large number of pixels
\citep{gould2013}. However, the model-fitting (statistical)
uncertainties only represent a portion of the true error budget for
the full-disk fits. In this case, the major source of systematic
uncertainty is likely to be the stellar mass profile. Pixels at
$r\gg\rg$ in the ALMA data dominate the full-disk fit, leading to very
tight constraints on $\Upsilon$ and consequently on \mbh\ as well. If
the assumed stellar mass profile shape is systematically wrong, model
fits will force \mbh\ to an incorrect (but ostensibly precise) value.

\subsection{Fits to smaller regions}

The stellar mass profile can be incorporated into the error budget on
\mbh\ if the uncertainty in its shape can be parameterized, with
addition of one or more free parameters to the model. Potential
sources of error in the mass profile shape include errors due to dust
extinction, errors in deprojection from two to three dimensions, and
$\Upsilon$ gradients due to stellar population age and/or metallicity
variations \citep{mcconnell2013}. Properly accounting for this range
of uncertainties over the entire radial range of the disk would be
challenging.

The alternative and simpler approach is to fit models over only
spatial pixels close to and within \rg, so that the BH accounts for a
larger fraction of the total enclosed mass. This makes the
model-fitting results less susceptible to errors due to the shape of
the stellar mass profile, and will appropriately broaden the
confidence range on \mbh\ since $\Upsilon$ will be less tightly
constrained.  Fits to smaller regions can also mitigate systematic
mismatch of flat-disk models to the disk's slightly warped
structure. Our goal is to fit models over the smallest possible
spatial region that permits successful fits with well-defined
confidence limits on \mbh\ (manifested by a smooth and regular curve
of \chisq\ vs.\ \mbh).

\begin{figure}
  \scalebox{0.4}{\includegraphics{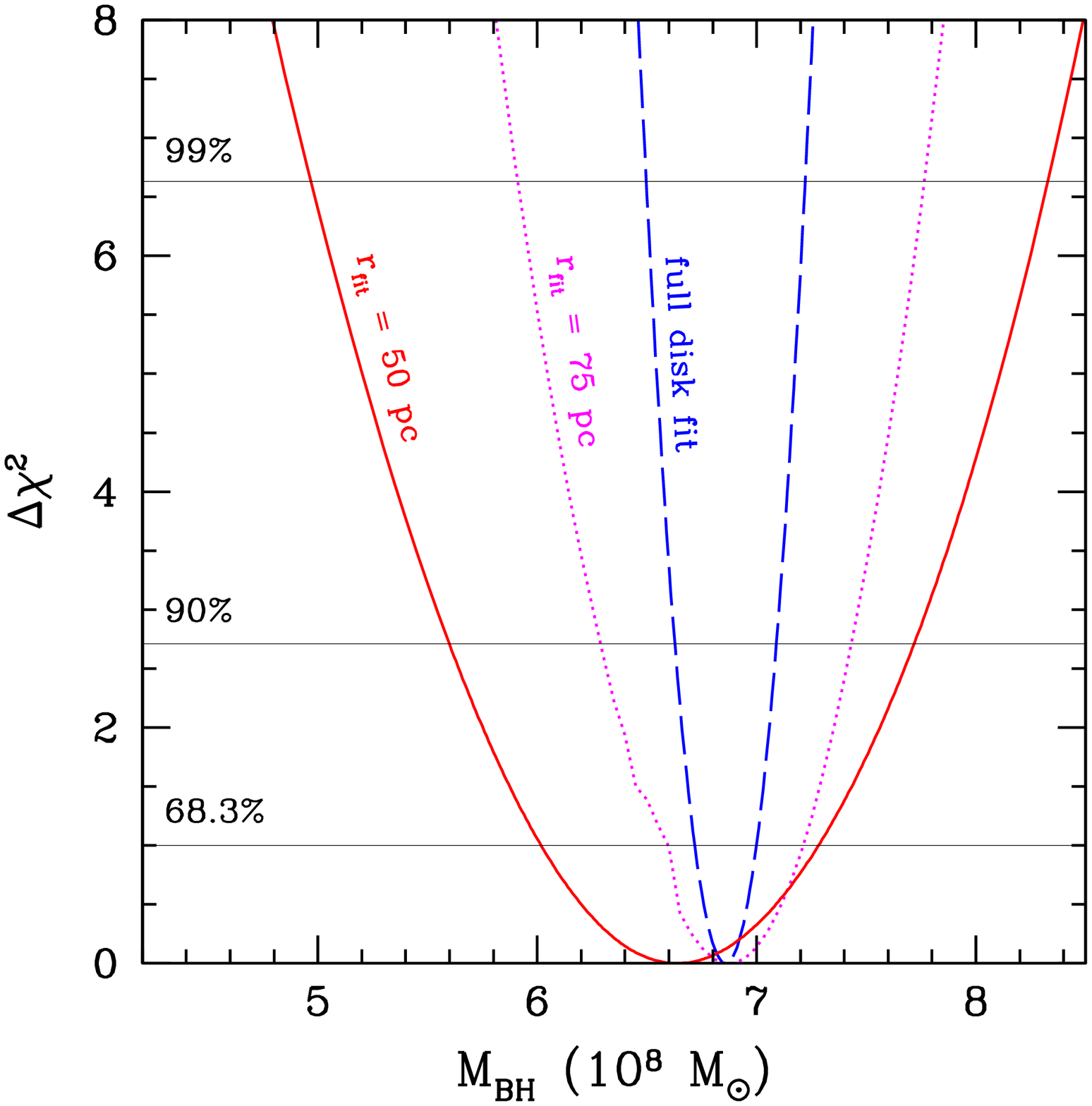}}
  \caption{Plot of $\Delta\chisq = \chisq - \chisq_\mathrm{min}$ for
    fits to the full disk ($\rfit=200$ pc), $\rfit=75$ pc with all
    parameters free (other than \mbh) at each trial value of \mbh, and
    $\rfit=50$ pc with $x_0$, $y_0$, and \vsys\ fixed. }
  \label{fig:chisq}
\end{figure}

We fit models to progressively smaller regions centered on the nucleus
and bounded by ellipses with semimajor and semiminor axis lengths of
\rfit\ and $\rfit\cos84\arcdeg$, for varying values of \rfit\ (where
the full disk corresponds to $\rfit=200$ pc). The curve of
\chisq\ vs.\ \mbh\ begins to show asymmetric structure for
$\rfit\leq75$ pc (Figure \ref{fig:chisq}), becoming increasingly
irregular and jagged for $\rfit\leq50$ pc due to the small number of
pixels and relatively low S/N in the fitting region. The irregular
structure in the \chisq\ curve is primarily the result of small
fluctuations in the best-fitting values of $x_0$, $y_0$, and
\vsys\ for different fixed values of \mbh. From the best-fitting model
with $\rfit=50$ pc, we take the values of these three parameters and
hold them fixed, calculating a grid of models with $\rfit=50$ pc for a
range of values of \mbh\ with the remaining five parameters free.

These constrained fits for $\rfit=50$ pc (containing 45 spatial
pixels) have 2515 degrees of freedom, and the best-fitting model has
$\chisq=2739.8$ and $\chisqdof=1.09$. The expected range on
\chisqdof\ for 2515 degrees of freedom is $1.00\pm0.03$, so this
result suggests that there is still a mild degree of systematic
model-data mismatch within $\rfit=50$ pc. The 68.3\% and 99\%
confidence ranges on \mbh\ are $(6.64_{-0.63}^{+0.65})\times10^8$ and
$(6.64_{-1.66}^{+1.69})\times10^8$ \msun, respectively. Other
parameters of the best-fitting model are $\Upsilon_R=7.83$ (giving
$\mstar = 1.8\times10^9$ \msun\ within $\rfit=50$ pc), $i=85\fdg2$,
$\Gamma=116\fdg7$, and $\sigmaturb=32.1$ \kms; the fixed value of
\vsys\ is 1562.2 \kms.  The error budget does not incorporate
uncertainty in the distance to NGC 1332; the derived \mbh\ scales
linearly with the assumed distance. The fits rule out $\mbh=0$ at
$>99.99\%$ confidence ($\Delta\chisq=74$ relative to the best-fitting
\mbh). We interpret the systematic changes in $i$ and $\Gamma$ for
smaller fitting regions as indications of the disk's warp; these
angles vary by only $\sim1\arcdeg$ from $\rfit=200$ pc down to
$\rfit=50$ pc or smaller regions.  This model implies $\rg = 25$ pc or
0\farcs23, determined as the radius within which $M_\star(\rg) =
\mbh$. Along the disk's minor axis, this projects to $\rg\cos i =
0\farcs02$.

For $\rfit<50$ pc, the \chisq\ curve again becomes irregular and
asymmetric, even with $x_0$, $y_0$, and \vsys\ held fixed. Achieving
stable fits for $\rfit<50$ pc is possible if the values of $i$,
$\Gamma$, $\sigmaturb$, and/or $f_0$ are fixed, but we choose
$\rfit=50$ pc for our final results since it is the smallest region
over which we can obtain successful fits with all of these parameters
free, and because \chisqdof\ increases slightly for values of
\rfit\ smaller than 50 pc.

\subsection{Turbulent velocity dispersion}
\label{sec:turbulence}

In Paper I, we carried out fits using a spatially uniform \sigmaturb,
and also ran models allowing for a central rise in \sigmaturb.  The
models allowing more flexibility in $\sigmaturb(r)$ strongly preferred
the presence of a steep increase in \sigmaturb\ at small radius,
reaching high (and likely unphysical) values of up to 270 \kms\ while
driving \mbh\ to zero. The fundamental problem was that the extreme
beam smearing due to the disk's high inclination produced very broad
line profiles near the disk center, and the model fits were unable to
distinguish between rapid rotation and turbulent broadening at small
radii.

The Cycle 3 data are much less susceptible to this particular
degeneracy, but with $\rg\cos i = 0\farcs02$ the minor-axis kinematics
are still unresolved by a factor of two, leading to significant beam
smearing of line profiles at small $r$. This blurring is evident in
the PVD as the ``fan'' of emission across a broad velocity range at
small radii (Figure \ref{fig:pvd}), and as the ``X''-shaped rise in
the \sigmalos\ map (Figure \ref{fig:momentmaps}). This velocity
structure is well matched by our models with uniform (and low) values
of \sigmaturb; a high central \sigmaturb\ is not required in order to
match the central increase in the observed \sigmalos.

In our best-fit model with $\rfit=50$ pc, for which $\sigmaturb=32.1$
\kms, the maximum value of $\sigmaturb/\vrot$ is 0.073, small enough
to justify our treatment of the disk as dynamically cold and
geometrically thin.  The $\rfit=50$ pc model suggests a modest central
increase in \sigmaturb\ relative to the outer disk, where our
full-disk fit found $\sigmaturb=22.2$ \kms. We carried out trial fits
to smaller regions of $\rfit=40$ or 30 pc and found that
\sigmaturb\ never exceeded 40 \kms\ for best-fitting models over any
fitting region. This demonstrates that the Cycle 3 data do not require
or prefer a high central \sigmaturb.

\section{Discussion and Conclusions}

This work presents the first BH mass derived from ALMA observations
that resolve the BH's gravitational radius of influence, demonstrating
ALMA's powerful capability to improve our understanding of local BH
demographics. The primary challenges for gas-dynamical measurements of
BH masses are identifying disks that exhibit suitably clean rotation,
and obtaining observations of sufficient angular resolution and S/N to
map the kinematics on the required scales. While a great deal of
effort has been invested in measurement of ionized gas kinematics,
primarily with \emph{Hubble Space Telescope} spectroscopic
observations, the cold molecular gas in circumnuclear disks is known
to be less turbulent than the ionized component, making molecular
kinematics a more accurate tracer of a galaxy's gravitational
potential than optical emission-line kinematics \citep{young2008,
  davis2013a}. Resolving \rg\ in CO observations of nearby galaxies
was at the limiting edge of the capabilities of earlier mm-wave
interferometers \citep{davis2013b}, but ALMA will be able to carry out
such observations for a large number of galaxies.

In Paper I we stressed the importance of resolving angular scales
corresponding to $\rg\cos i$ (the BH radius of influence projected
along the disk's minor axis) in order to fully resolve disk kinematics
within the BH's sphere of influence. This Cycle 3 observation of NGC
1332 fully resolves \rg, but $\rg\cos i$ is unresolved by a factor of
two. Thus, the observed kinematics are still blurred by beam smearing
in the minor-axis direction, as seen in the \sigmalos\ map and the
PVD, but this effect is far less severe than in our Cycle 2
observation. Higher resolution observations would be required in order
to determine the disk's central $\sigmaturb(r)$ profile accurately and
reduce the remaining degeneracy between rotation and turbulence in the
inner disk. This would permit even tighter constraints on \mbh\ and
provide unique information on the physical properties of molecular
gas deep within the gravitational potential well of a supermassive
BH.

The ideal situation for a gas-dynamical BH mass measurement is
achieved when observations have sufficiently high S/N and angular
resolution that \mbh\ can be well constrained by models fitted only to
points within $r<\rg$, so that stellar mass represents a small
contribution to the total enclosed mass and systematic uncertainties
in the stellar mass profile have a negligible impact on the
measurement. This has previously been accomplished for a few objects
using ionized gas kinematics \citep[e.g, M87;][]{macchetto1997,
  walsh2013} as well as for H$_2$O maser disk galaxies such as NGC
4258 \citep{miyoshi1995, kuo2011}. ALMA is certainly capable of
achieving this goal for observations of CO or other molecular
tracers. With an optimal combination of high angular resolution and
S/N, ALMA observations of other targets could in principle provide BH
mass precision at the level of a few percent in future measurements.
All else being equal, the most precise BH masses will be determined
from observations of disks at intermediate inclination angles
satisfying the competing requirements that $\rg\cos i$ is resolved and
$v\sin i$ is large enough to permit accurate kinematic mapping. NGC
1332 is a challenging target in that the high disk inclination
requires observations at very high angular resolution. At high
resolution the S/N per spatial pixel is low, and for this measurement,
S/N is the primary limiting factor for determining \mbh. Significant
improvement in S/N would require a very large investment of ALMA time,
however, far more than the $\sim5$ hours (including calibrations and
overhead time) used for this observation.

The CO kinematics are highly inconsistent with a central mass as high
as $\mbh=(1.45\pm0.20)\times10^9$ \msun, the value found by
\citet{rusli2011} based on stellar-dynamical modeling of VLT
adaptive-optics spectroscopic data. For our constrained fits with
$\rfit=50$ pc, we find $\Delta\chisq=101$ at $\mbh=1.45\times10^9$
\msun, and $\Delta\chisq=67$ for $\mbh=1.25\times10^9$ \msun.  This
disagreement is particularly stark in that we are using the same
stellar luminosity profile measured by \citet{rusli2011} in our
dynamical models, and our fits find a stellar mass-to-light ratio very
similar to their best-fit value. It would be worthwhile to revisit the
stellar-dynamical models for NGC 1332 to determine whether this
disagreement can be resolved.

Our best-fitting BH mass is consistent with the value
$5.2_{-2.8}^{+4.1}\times10^8$ \msun\ (90\% confidence) found by
\citet{humphrey2009} from modeling the hydrostatic equilibrium of the
diffuse X-ray emitting hot gas halo of NGC 1332. This agreement
provides new motivation to explore the hydrostatic equilibrium method
as a promising approach for measurement of the central mass profiles
in elliptical galaxies and to carry out additional direct comparisons
with other mass measurement techniques.

The scatter in the \msigma\ correlation at low redshift provides a
fossil record of the cosmic history of BH evolution through accretion
and mergers \citep{robertson2006, peng2007}. Determination of this
scatter requires a large sample of \mbh\ measurements with reliable
measurement uncertainties.  The factor of two disagreement between our
gas-dynamical \mbh\ value and the stellar-dynamical result from
\citet{rusli2011} serves as an important cautionary note regarding the
measurement and interpretation of the scatter in the
\msigma\ relation, which is primarily determined from
stellar-dynamical measurements \citep{gultekin2009, mcconnellma,
  kormendyho}. Our results suggest that much more effort will be
required in order to arrive at a definitive empirical understanding of
the scatter in the local BH-bulge correlations. ALMA is poised to make
a major contribution to our understanding of local BH demographics, if
significant numbers of cleanly rotating molecular disks similar to the
NGC 1332 disk can be identified.

\acknowledgements

This paper makes use of data from ALMA program 2015.1.00896.S. ALMA is
a partnership of ESO (representing its member states), NSF (USA) and
NINS (Japan), together with NRC (Canada) and NSC and ASIAA (Taiwan),
in cooperation with the Republic of Chile. The Joint ALMA Observatory
is operated by ESO, AUI/NRAO and NAOJ. The National Radio Astronomy
Observatory is a facility of the National Science Foundation operated
under cooperative agreement by Associated Universities, Inc.  We thank
Jens Thomas for providing the stellar mass profile from
\citet{rusli2011}.

LCH acknowledges support from the Chinese Academy of Science through
grant XDB09030102 (Emergence of Cosmological Structures) from the
Strategic Priority Research Program, and from the National Natural
Science Foundation of China through grant 11473002.

\emph{Facilities:} \facility{ALMA}


\begin{thebibliography}{}

\bibitem[Barth et al.(2016)]{barth2016} Barth, A. J., Darling, J.,
  Baker, A. J., Boizelle, B. D., Buote, D. A., Ho, L. C., \& Walsh,
  J. L. 2016, \apj, in press (arXiv:1603.04523).

\bibitem[Davis et al.(2013a)]{davis2013a} Davis, T.~A., Alatalo, 
K., Bureau, M., et al.\ 2013a, \mnras, 429, 534 
  
\bibitem[Davis et al.(2013b)]{davis2013b} Davis, T.~A., Bureau, M., 
Cappellari, M., Sarzi, M., \& Blitz, L.\ 2013b, \nat, 494, 328 

\bibitem[Gould(2013)]{gould2013} Gould, A.\ 2013, arXiv:1303.0834 

\bibitem[G{\"u}ltekin et al.(2009)]{gultekin2009} G{\"u}ltekin, K., 
Richstone, D.~O., Gebhardt, K., et al.\ 2009, \apj, 698, 198 
  
\bibitem[Humphrey et al.(2009)]{humphrey2009} Humphrey, P.~J., 
Buote, D.~A., Brighenti, F., Gebhardt, K., 
\& Mathews, W.~G.\ 2009, \apj, 703, 1257 

\bibitem[Kormendy \& Ho(2013)]{kormendyho} Kormendy, J., \& Ho,
  L.~C.\ 2013, \araa, 51, 511

\bibitem[Kuo et al.(2011)]{kuo2011} Kuo, C.~Y., Braatz, J.~A., 
Condon, J.~J., et al.\ 2011, \apj, 727, 20 

\bibitem[Macchetto et al.(1997)]{macchetto1997} Macchetto, F.,
  Marconi, A., Axon, D.~J., et al.\ 1997, \apj, 489, 579

\bibitem[McConnell et al.(2013)]{mcconnell2013} McConnell, N.~J., 
Chen, S.-F.~S., Ma, C.-P., et al.\ 2013, \apjl, 768, L21 

\bibitem[McConnell \& Ma(2013)]{mcconnellma} McConnell, N.~J.,
  \& Ma, C.-P.\ 2013, \apj, 764, 184
  
\bibitem[McMullin et al.(2007)]{mcmullin2007} McMullin, J.~P., Waters,
  B., Schiebel, D., Young, W., \& Golap, K.\ 2007, Astronomical Data
  Analysis Software and Systems XVI, 376, 127

\bibitem[Miyoshi et al.(1995)]{miyoshi1995} Miyoshi, M., Moran, J., 
Herrnstein, J., et al.\ 1995, \nat, 373, 127 
  
\bibitem[Onishi et al.(2015)]{onishi2015} Onishi, K., Iguchi, S., 
Sheth, K., \& Kohno, K.\ 2015, \apj, 806, 39 

\bibitem[Peng(2007)]{peng2007} Peng, C.~Y.\ 2007, \apj, 671, 
1098 

\bibitem[Robertson et al.(2006)]{robertson2006} Robertson, B., 
Hernquist, L., Cox, T.~J., et al.\ 2006, \apj, 641, 90 

\bibitem[Rusli et al.(2011)]{rusli2011} Rusli, S.~P., Thomas, J., 
Erwin, P., et al.\ 2011, \mnras, 410, 1223 

\bibitem[Walsh et al.(2013)]{walsh2013} Walsh, J.~L., Barth, 
A.~J., Ho, L.~C., \& Sarzi, M.\ 2013, \apj, 770, 86 

\bibitem[Young et al.(2008)]{young2008} Young, L.~M., Bureau, M., 
\& Cappellari, M.\ 2008, \apj, 676, 317-334 





\end{thebibliography}
\end{document}